\documentclass[11pt,a4paper]{article}

\pdfoutput=1

\usepackage{jcappub}
\usepackage{graphicx}
\usepackage{float}
\usepackage{slashed}
\usepackage[font=small, labelfont=bf]{caption}
\usepackage[labelformat=simple]{subcaption}

\usepackage{amsmath, amsthm, amssymb, amsfonts}
\usepackage{multirow}
\usepackage[numbers,sort&compress]{natbib}
\usepackage{hyperref}

\def\beq{\begin{align}}
\def\eeq{\end{align}}

\newcommand{\vo}{\mathcal{V}}

\newcommand{\bi}{\begin{itemize}}
\newcommand{\ei}{\end{itemize}}
\newcommand{\ben}{\begin{enumerate}}
\newcommand{\een}{\end{enumerate}}
\newcommand{\be}{\begin{equation}}
\newcommand{\ee}{\end{equation}}

\newcommand{\F}{\mathcal{F}}
\newcommand{\comments}[1]{}
\def\nn{\nonumber}

\def\PBH{{\scriptscriptstyle \rm PBH}}
\def\BH{{\scriptscriptstyle \rm BH}}
\def\CMB{{\scriptscriptstyle \rm CMB}}

\def\KK{{\scriptscriptstyle \rm KK}}
\def\W{{\scriptscriptstyle \rm W}}
\def\DM{{\scriptscriptstyle \rm DM}}

\def\vo{{{\cal{V}}}}

\newcommand{\mc}{\mathcal}

\newcommand{\beqa}{\begin{eqnarray}}
\newcommand{\eeqa}{\end{eqnarray}}

\providecommand{\bea}{\begin{eqnarray}}
 
 \providecommand{\rm}{\mathrm}
\providecommand{\eea}{\end{eqnarray}}

\def\M{{\scriptscriptstyle {\rm M}}}

\def\K3{{\scriptscriptstyle {\rm K3}}}
\def\dP{{\scriptscriptstyle {\rm dP}}}
\def\up{{\scriptscriptstyle {\rm up}}}
\def\F4{{\scriptscriptstyle {\rm F}^4}}

\def\46{{\scriptscriptstyle {\rm 4-6}}}
\def\24{{\scriptscriptstyle {\rm 2-4}}}

\title{Primordial Black Holes from String Inflation}

\author[a,b,c]{Michele Cicoli,}
\author[a,b]{Victor A. Diaz,}
\author[a,b]{Francisco G. Pedro}

\affiliation[a]{\small Dipartimento di Fisica e Astronomia, Universit\`a di Bologna, \\ via Irnerio 46, 40126 Bologna, Italy}
\affiliation[b]{\small INFN, Sezione di Bologna, viale Berti Pichat 6/2, 40127 Bologna, Italy}
\affiliation[c]{\small Abdus Salam ICTP, Strada Costiera 11, Trieste 34151, Italy}

\emailAdd{mcicoli@ictp.it}
\emailAdd{diaz@bo.infn.it}
\emailAdd{francisco.pedro@bo.infn.it}

\abstract{We present a single-field string inflationary model which allows for the generation of primordial black holes in the low mass region where they can account for a significant fraction of the dark matter abundance. The potential is typical of type IIB fibre inflation models and features a plateau at CMB scales and a near inflection point at large distance scales where the power spectrum is enhanced due to a period of ultra slow-roll. The tunability of the underlying parameters is guaranteed by scanning through the string landscape and their stability against quantum corrections is ensured by an effective shift symmetry.}


\begin{document}
\maketitle

\section{Introduction}
\label{Intro}

The origin of dark matter remains one of the biggest mysteries in fundamental physics. One of the simplest explanations, which would rely  neither on the presence of new particles nor on modifications of the gravitational interaction, is black holes. An interesting region in parameter space where the contribution of black holes to the total dark matter abundance could be between $10\%$ and $100\%$ depending on astrophysical uncertainties is $10^{-17} M_\odot \lesssim M_\BH\lesssim 10^{-13} M_\odot$ \cite{Sasaki:2018dmp}, where the lower bound comes from extra-galatic $\gamma$ rays produced due to Hawking evaporation \cite{Hawking:1974rv}. This region, even if it is far from the one probed by LIGO, is very interesting since there is no known astrophysical explanation for black hole formation in this small mass window.\footnote{Depending on the interpretation of astrophysical and cosmological data, X-ray and CMB observations seem to rule out the case where black holes in the LIGO mass region can constitute a fraction of the dark matter abundance above $10\%$ \cite{Ali-Haimoud:2016mbv, Poulin:2017bwe, Gaggero:2016dpq}. Moreover the single-field inflationary dynamics seems to be very unlikely to generate black holes with masses as large as a few solar masses when the scalar spectral index is required to be compatible with CMB data.} On the other hand, these tiny black holes could be seeded by the dynamics of the early universe \cite{Hawking:1971ei, Carr:1974nx}. A tantalizing idea for the formation of these primordial black holes (PBHs) relies on an amplification of the density perturbations during inflation of order $\delta\rho \sim 0.1 \,\rho$ which then collapse to form PBHs at horizon re-entry. 

This enhancement of the scalar power spectrum has to take place at momentum scales which are much larger than the ones associated with CMB observations where $\delta\rho \sim 10^{-5} \rho$. From the theoretical point of view, it is therefore important to identify mechanisms to generate the necessary enhancement at the right scales. Guided again by simplicity, we focus on single field inflationary models which also reproduce the Planck data rather well \cite{Ade:2015lrj}.\footnote{For PBH formation in multi-field inflationary models see \cite{Kawasaki:2012wr,Carr:2016drx,Garcia-Bellido:2016dkw}.} It has already been pointed out that the required inflationary potentials feature a slow-roll behaviour followed by a near inflection point region where the power spectrum is amplified since the system enters an ultra slow-roll regime \cite{Ivanov:1994pa, Garcia-Bellido:2017mdw, Ezquiaga:2017fvi, Germani:2017bcs, Ballesteros:2017fsr}.

Despite the fact that dark matter as PBHs formed during single-field inflation might seem a very appealing idea, its explicit realisation in concrete models has turned out to be rather complicated since the inflationary potential has to possess enough tuning freedom to allow for such  dynamics \cite{Hertzberg:2017dkh}. Examples based on a radiative plateau have been recently studied in \cite{Ezquiaga:2017fvi, Germani:2017bcs, Ballesteros:2017fsr}. This is a bottom up perspective which tries to single out the simplest potential which allows for PBH formation via an inflationary plateau followed by a near inflection point. However this approach ignores the fundamental issue of deriving the model from a UV consistent theory  where a symmetry argument can protect the flatness of the potential against quantum corrections.

In this paper we shall instead take a more top down approach and search for concrete examples of inflationary models in string theory whose structure is rich enough to allow for PBH formation. One of the main advantages of embedding inflation in string theory is the possibility to motivate the presence of a symmetry which can protect the inflaton potential against quantum corrections which can spoil its flatness \cite{Baumann:2009ni, Burgess:2013sla}. Particularly interesting cases include inflaton candidates which are pseudo Nambu-Goldstone bosons associated with slightly broken shift symmetries. Abelian symmetries involves both axions \cite{Pajer:2013fsa}, which are associated to compact $U(1)$ factors, and K\"ahler moduli \cite{Cicoli:2011zz}, which are associated with non-compact rescaling symmetries \cite{Burgess:2014tja}.\footnote{The non-Abelian case leads to a multi-field inflationary scenario which tends to be disfavoured by non-Gaussianity observations \cite{Ade:2015lrj}.}

This global rescaling symmetry is explicitly realised at tree-level in type IIB no-scale models since the K\"ahler moduli $\tau$ remain exact flat directions but needs to be slightly broken to generate the inflationary potential. This can be done either by non-perturbative effects or by perturbative power-law corrections which become exponential in terms of the canonically normalised inflaton: $V_0/\tau^n \sim V_0 \,e^{-n \phi/f}$. Notice that the shape of the inflationary potential is determined by both the effective `decay constant' $f$, i.e. the geometry of the moduli space (determined by the topology of the divisor whose volume is parameterised by the inflaton) and $n$, i.e. the exact moduli-dependence of the symmetry-breaking effects which develop the inflationary potential \cite{Burgess:2016owb}. Once a proper uplifting to dS has been achieved via the addition of a constant contribution (which can have several dynamical origins \cite{Kachru:2003aw, Kallosh:2014wsa, Bergshoeff:2015jxa, Burgess:2003ic, Cicoli:2013cha, Cicoli:2015ylx, Cicoli:2012fh}), these models tend to give rise to an inflationary potential of the schematic form \cite{Burgess:2016owb}:
\be
V_{\rm inf} = V_0 \left(1- \,e^{-n \phi/f}\right)\,.
\label{Vinf1}
\ee
These models go under the name of \textit{Fibre Inflation} since the underlying Calabi-Yau compactification manifold has a typical fibration structure \cite{Cicoli:2008gp, Broy:2015zba, Cicoli:2016chb}. They are interesting since they drive inflation successfully via a plateau-like region at large $\phi$ and also allow  for a detailed analysis of the post-inflationary evolution \cite{Cicoli:2010ha, Cabella:2017zsa, Antusch:2017flz}. Moreover they provide string theory embeddings of Starobinsky inflation \cite{Starobinsky:1980te} and supergravity $\alpha$-attractors \cite{Kallosh:2013maa, Kallosh:2015zsa, Carrasco:2015pla} (where in our notation $\alpha \simeq (f/n)^2$). Nevertheless the potential (\ref{Vinf1}) is too simple to generate PHBs via a period of ultra slow-roll dynamics towards the end of inflation. However recent global constructions of fibre inflation models in concrete Calabi-Yau orientifolds with explicit brane setup and closed string moduli stabilisation have revealed the existence of new string loop corrections which look schematically like \cite{Cicoli:2016xae, Cicoli:2017axo}:
\be
\delta V_{\rm inf} = - \epsilon_1 V_0 \,\frac{e^{2 n \phi/f}}{1+ \epsilon_2\,e^{3 n \phi/f}} \,,
\label{Vinf2}
\ee
where $\epsilon_1 \ll 1$ and $\epsilon_2 \ll 1$ are two parameters which are tunable since they depend on background fluxes and the Calabi-Yau intersection numbers, and turn out to be naturally small since they are suppressed by inverse powers of the compactification volume, an exponentially large quantity \cite{Balasubramanian:2005zx, Cicoli:2008va}. Thanks to the additional perturbative contribution (\ref{Vinf2}), we will show that fibre inflation models are rich enough to produce a near inflation point region before the end of inflation which is perfectly suitable to generate PBHs in the mass window $10^{-17} M_\odot \lesssim M_\PBH\lesssim 10^{-13} M_\odot$ where they could constitute a significant fraction of the total dark matter abundance. 

As pointed out in \cite{Kannike:2017bxn, Germani:2017bcs, Ballesteros:2017fsr, Motohashi:2017kbs}, the slow-roll approximation ceases to be valid in the near inflection point region. The primordial power spectrum has to be computed by solving the Mukhanov-Sasaki equations for the curvature perturbations \cite{Sasaki:1986hm, Mukhanov:1988jd}. By following this procedure, we shall show that the primordial power spectrum can feature the required enhancement for appropriate values of the underlying parameters. Let us stress that even if the choice of microscopic parameters needed for successful PBH formation looks very non-generic from the string landscape point of view, the values of these parameters are technically natural since they are protected against large quantum corrections by the effective rescaling shift symmetry typical of these models \cite{Burgess:2014tja}.

This paper is organised as follows. In Sec. \ref{FIReview} we provide a very brief review of fibre inflation models while in Sec. \ref{sec:PBH} we describe the mechanism of PBH generation in some detail. In Sec. \ref{PBHFibre} we then perform a careful analysis of the process of PBH formation in fibre inflation by implementing the Mukhanov-Sasaki formalism to derive the primordial power spectrum. We finally discuss our results and present our conclusions in Sec. \ref{Concl}.

\section{Fibre inflation models}
\label{FIReview}

Fibre inflation is a class of string inflationary models built within the framework of type IIB flux compactifications \cite{Cicoli:2008gp, Broy:2015zba, Cicoli:2016chb, Burgess:2016owb}. The inflaton $\tau_\K3$ is a K\"ahler modulus controlling the size of a K3 divisor fibred over a $\mathbb{P}^1$ base with volume $t_{\mathbb{P}^1}$. The simplest fibre inflation models feature a Calabi-Yau (CY) volume which looks like:
\be
\vo = t_{\small \mathbb{P}^1}\tau_\K3 - \tau_\dP^{3/2}\,,
\ee
where $\tau_\dP$ is the volume of a diagonal del Pezzo divisor which supports non-perturbative effects. Several effects come into play to stabilise the K\"ahler moduli in a typical LVS vacuum \cite{Balasubramanian:2005zx, Cicoli:2008va}. At leading order in a $1/\vo\ll 1$ expansion only two directions, $\vo$ and $\tau_\dP$, are lifted by non-perturbative contributions to the superpotential $W$ \cite{Kachru:2003aw} and perturbative $\alpha'$ corrections to the K\"ahler potential $K$ \cite{Becker:2002nn, Minasian:2015bxa, Bonetti:2016dqh}.\footnote{At this level of approximation, also the axionic partner of $\tau_\dP$ is fixed by non-perturbative effects.} Hence the remaining flat direction, which can be parametrised by $\tau_\K3$, represents a very promising inflaton candidate since it enjoys an effective non-compact rescaling symmetry which can be used to protect the flatness of the inflationary potential against quantum corrections \cite{Burgess:2014tja}.\footnote{There are actually other two flat directions corresponding to the axions associated with the base and the fibre which turn out to be much lighter than $\tau_\K3$. Hence these fields acquire isocurvature fluctuations during inflation. However present strong bounds on isocurvature fluctuations do not apply to our case since these axions tend to be too light to behave as dark matter.} 

In order to generate the inflationary potential, this effective shift symmetry has to be slightly broken. This is realised by open string 1-loops which depend on all K\"ahler moduli \cite{Berg:2005ja, Berg:2007wt, Berg:2014ama, Haack:2015pbv} but are subdominant with respect to the leading $\alpha'$ effects thanks to the extended no-scale structure typical of these models \cite{Cicoli:2007xp}. Other contributions to the inflationary potential arise from higher derivative $\alpha'$ effects \cite{Ciupke:2015msa, Grimm:2017okk} but these are also $\vo$-suppressed if the superspace derivative expansion is under control \cite{Cicoli:2013swa}. Moreover, all these corrections give rise to an AdS vacuum which needs to be uplifted to dS by the inclusion of anti-branes \cite{Kachru:2003aw, Kallosh:2014wsa, Bergshoeff:2015jxa}, hidden sector T-branes \cite{Burgess:2003ic, Cicoli:2013cha, Cicoli:2015ylx} or non-perturbative effects at singularities \cite{Cicoli:2012fh}. It is important to stress that all these uplifting effects are inflaton-independent since they depend just on the overall volume $\vo$. Thus they give rise to a constant contribution to the inflationary potential which is crucial to develop a plateau-like behaviour at large inflaton values. 

After canonical normalisation of the inflaton field, the resulting potential is qualitatively very similar to the one of Starobinsky inflation \cite{Starobinsky:1980te} and $\alpha$-attractor supergravity models \cite{Kallosh:2013maa, Kallosh:2015zsa, Carrasco:2015pla}. In fact, fibre inflation models require a trans-Planckian field range to obtain enough e-foldings of inflationary expansion, and so they can predict a tensor-to-scalar ratio as large as $r\sim 0.005 - 0.01$. These models are particularly interesting also because they can be embedded into globally consistent CY orientifold compactifications with an explicit brane setup and chiral matter \cite{Cicoli:2016xae, Cicoli:2017axo}. In the study of concrete CY realisations of string models where the inflaton is a K\"ahler modulus, it has been recently realised that the underlying K\"ahler cone conditions set strong geometrical constraints on the allowed inflaton range \cite{Cicoli:2018tcq}. Interestingly, it has been found that the distance travelled by inflaton in field space can generically be trans-Planckian only for K3-fibred CY threefolds which are exactly the necessary ingredients to construct fibre inflation models.

The two moduli which are stabilised at leading order in $1/\vo$ are heavier than the Hubble constant whose size is set by the uplifting contribution. Hence $\vo$ and $\tau_\dP$ do not play a significant r\^ole during inflation which is instead driven mainly by the light field $\tau_\K3$. Fibre inflation models are therefore, to a very good level of approximation, single-field inflationary models whose potential looks like \cite{Cicoli:2008gp, Broy:2015zba, Cicoli:2016chb}:
\be
V_{\rm inf}=\left(\frac{C_\up}{\vo^{4/3}} + g_s^2 \,\frac{C_\KK}{\tau_\K3^2} + \frac{W_0^2}{\sqrt{g_s}}\frac{\epsilon_\F4}{\vo\,\tau_\K3}  -\frac{C_\W}{\vo\sqrt{\tau_\K3}} +g_s^2\,D_\KK\,\frac{\tau_\K3}{\vo^2} +\delta_\F4\,\frac{W_0^2}{\sqrt{g_s}}\,\frac{\sqrt{\tau_\K3}}{\vo^2}\right)\frac{W_0^2}{\vo^2}\,,
\label{PotFI}
\ee
where $g_s\ll 1$ is the string coupling and $W_0\sim \mc{O}(1-10)$ is the superpotential generated by background fluxes which is constant after the dilaton and the complex structure moduli are stabilised at tree-level. $C_\up$ controls the uplifting contribution and, depending on the particular mechanism employed it can have a different dependence on the internal volume $\vo$, background or gauge fluxes. $C_\KK>0$, $D_\KK>0$ and $C_\W$ are the coefficients of 1-loop open string corrections which come respectively from the tree-level exchange of closed Kaluza-Klein strings between non-intersecting stacks of branes, and winding closed strings between intersecting branes \cite{Berg:2005ja, Berg:2007wt, Berg:2014ama, Haack:2015pbv}. These constants are also functions of the the vacuum expectation values of the complex structure moduli and are expected to be of order unity: $C_\KK\sim D_\KK\sim C_\W \sim \mc{O}(1)$. On the other hand, $\epsilon_\F4$ and $\delta_\F4$ are the coefficients of higher derivative $\alpha'$ $F^4$ effects which depend just on the topological properties of the underlying geometry and are expected to be positive but relatively small: $\epsilon_\F4\sim\delta_\F4\sim \mc{O}(10^{-3})$ \cite{Ciupke:2015msa, Grimm:2017okk}. 

The potential (\ref{PotFI}) is rich enough to generate a minimum for small $\tau_\K3$, an inflationary plateau-like behaviour at large $\tau_\K3$ and finally a steepening region at very large $\tau_\K3$ where the system is in a fast-roll regime.\footnote{Pre-inflationary fast to slow-roll transitions in fibre inflation models can give rise to a power loss at large angular scales \cite{Cicoli:2013oba, Pedro:2013pba, Cicoli:2014bja}.} In order to perform a proper study of the inflationary dynamics, the field $\tau_\K3$ has to be written in terms of its canonically normalised counterpart $\phi$ as \cite{Cicoli:2008gp}:
\be
\tau_\K3 = e^{\frac{2}{\sqrt{3}} \phi} = \langle\tau_\K3\rangle \,e^{\frac{2}{\sqrt{3}} \hat\phi}\,,
\label{CanNorm}
\ee
where we have expanded $\phi$ around its minimum as $\phi=\frac{\sqrt{3}}{2} \ln\langle\tau_\K3\rangle + \hat\phi$. Substituting (\ref{CanNorm}) in (\ref{PotFI}), we end up with:
\be
V_{\rm inf}=V_0 \left(C_1 + C_2 \,e^{-\frac{4}{\sqrt{3}} \hat\phi} + C_3\,e^{-\frac{2}{\sqrt{3}} \hat\phi}  - \,e^{-\frac{1}{\sqrt{3}} \hat\phi} 
+C_4\,e^{\frac{2}{\sqrt{3}} \hat\phi} +C_5\,e^{\frac{1}{\sqrt{3}} \hat\phi}\right)\,,
\label{PotFInorm}
\ee
where, parameterising the inflaton minimum as $\langle\tau_\K3\rangle^{3/2} \equiv \gamma\, \vo$, we have:
\bea
V_0 &=&  \frac{C_\W\,W_0^2}{\gamma^{1/3}\vo^{10/3}}\,, \qquad
C_1 = \gamma^{1/3}\,\frac{C_\up}{C_\W}\,, \qquad
C_2 = g_s^2 \,\frac{C_\KK}{\gamma\,C_\W}\,, \nn \\
C_3 &=& \frac{W_0^2}{\gamma^{1/3}\,C_\W\,\sqrt{g_s}}\frac{\epsilon_\F4}{\vo^{1/3}}\,, \qquad
C_4 = \gamma\,g_s^2\,\frac{D_\KK}{C_\W}\,, \qquad C_5 = \gamma\,C_3\,\frac{\delta_\F4}{\epsilon_\F4}\,. 
\label{Param}
\eea
The potential (\ref{PotFInorm}) can have a plateau-like region which can support enough efoldings of inflation only if the coefficients of the positive exponentials are suppressed, i.e. $C_4\ll 1$ and $C_5\ll 1$, which, in turn, requires $\gamma \ll 1$. This is naturally achieved if the three negative exponentials compete to give a minimum since this can happen when $\gamma \sim g_s^2 \ll 1$. The inflationary plateau is then generated mainly by the fourth term in (\ref{PotFInorm}). Notice that the Hubble constant during inflation is set by $V_0$ and scales as $H^2\sim M_p^2 /\vo^{10/3}$.\footnote{$M_p$ denotes the reduced Planck mass $M_p = 1/\sqrt{8\pi G } \simeq 2.4\cdot 10^{18}$ GeV.} The mass of the inflaton around the minimum is of order $H$ but then quickly becomes exponentially smaller than $H$ for $\hat\phi>0$. 

Even if (\ref{PotFInorm}) is a very promising potential to drive inflation, it is not rich enough to generate primordial black holes due to the requirement of a  significant enhancement of the power spectrum at large momentum scales. However recent explicit constructions of fibre inflation models in concrete type IIB CY compactifications with D3/D7-branes and O3/O7-planes have reproduced the potential (\ref{PotFI}) in a slightly generalised form since \cite{Cicoli:2016xae, Cicoli:2017axo}:
\bi
\item In general the coefficient $C_\W$ is not a constant but a function of the fibre modulus $\tau_\K3$ of the form:
\be
C_\W \quad\to\quad C_\W(\tau_\K3) = C_\W - \frac{A_\W\sqrt{\tau_\K3}}{\sqrt{\tau_\K3}-B_\W}\,,
\label{Cnew}
\ee
where the parameters $C_\W\sim\mc{O}(1)$ and $A_\W\sim\mc{O}(1)$ depend on the vacuum expectation values of the complex structure moduli, while $B_\W\sim\mc{O}(1)$ depends on topological properties of the underlying CY threefold like the intersection numbers and the Euler number.

\item The effective action features additional winding 1-loop corrections to the inflationary potential which will turn out to be crucial for the formation of primordial black holes and look like:
\be
\delta V_\W = W_0^2\,\frac{\tau_\K3}{\vo^4}
\left(D_\W - \frac{G_\W}{1+R_\W \,\frac{\tau_\K3^{3/2}}{\vo}}\right)\,,
\label{AddPot}
\ee
where again $D_\W\sim\mc{O}(1)$ and $G_\W\sim\mc{O}(1)$ become constants only after complex structure moduli stabilisation, while $R_\W\sim\mc{O}(1)$ depends on the topological features of the extra dimensions.
\ei
Depending on the details of a given brane setup (in particular the presence of intersections between D-branes and O-planes and the topological properties of two-cycles where different stacks can intersect), several contributions to the generic scalar potential (\ref{PotFI}), supplemented with (\ref{Cnew}) and (\ref{AddPot}), can be absent by construction. In what follows, we shall therefore focus just on winding 1-loop corrections that represent the simplest situation which can lead to a successful generation of primordial black holes. This is justified for example by the fact that the global chiral embedding of fibre inflation presented in \cite{Cicoli:2017axo} does not feature any Kaluza-Klein loop correction, i.e. $C_\KK=D_\KK=0$.\footnote{Even if both $C_\KK$ and $D_\KK$ are non-zero, in a vast region of the parameter space, Kaluza-Klein loops would still be subdominant with respect to winding loops due to the extra factors of $g_s^2\ll 1$ in (\ref{Param}). This is due to the fact that Kaluza-Klein loops feature an extended no-scale cancellation, and so they contribute to the scalar potential effectively only at 2-loop order \cite{Cicoli:2007xp}.} Moreover higher derivative $F^4$ terms tend also to be negligible since, as can be seen from (\ref{Param}), they should be suppressed by both inverse volume powers and by $\epsilon_\F4\ll 1$ and $\delta_\F4\ll 1$. Hence in Sec. \ref{PBHFibre} we shall study primordial black hole formation for the following simplified inflationary potential:
\be
V_{\rm inf}= \frac{W_0^2}{\vo^3}\left[\frac{C_\up}{\vo^{1/3}}  -\frac{C_\W}{\sqrt{\tau_\K3}} + \frac{A_\W}{\sqrt{\tau_\K3}-B_\W}
 +\frac{\tau_\K3}{\vo} \left(D_\W-\frac{G_\W}{1+R_\W \,\frac{\tau_\K3^{3/2}}{\vo}}\right)\right]\,,
 \label{eq:Vinf}
\ee
which, when expressed in terms of the canonically normalised inflaton shifted from its minimum, takes the form:
\be
V_{\rm inf}=V_0 \left[C_1  - \,e^{-\frac{1}{\sqrt{3}} \hat\phi} \left(1-\frac{C_6}{1-C_7\,e^{-\frac{1}{\sqrt{3}} \hat\phi}}\right)
+C_8\,e^{\frac{2}{\sqrt{3}} \hat\phi} \left(1- \frac{C_9}{1+C_{10}\,e^{\sqrt{3} \hat\phi}}\right)\right]\,,
\label{PotNorm}
\ee
with:
\bea
C_6 &=&  \frac{A_\W}{C_\W}\sim \mc{O}(1)\,, \qquad
C_7 = \,\frac{B_\W}{\gamma^{1/3}\vo^{1/3}}\sim \mc{O}(1)\,, \qquad
C_8 = \gamma\, \frac{D_\W}{C_\W}\ll 1 \,, \nn \\
C_9 &=& \frac{G_\W}{D_\W}\sim \mc{O}(1)\,, \qquad
C_{10} = \gamma\,R_\W \ll 1\,. 
\label{NewParam}
\eea

\section{PBH formation}
\label{sec:PBH}

Primordial black holes form when large and relatively rare density perturbations re-enter the Hubble horizon and undergo gravitational collapse. The fraction of the total energy density in PBHs with mass $M$ at PBH formation is given by:
\be
\beta_{\rm f}(M)=\frac{\rho_\PBH (M)}{\rho_{\rm tot}}\Big|_{\rm f}\,.
\label{betaf}
\ee
The curvature perturbations are assumed to follow a Gaussian distribution with width $\sigma_\M\equiv\sigma(M)$.\footnote{See \cite{Franciolini:2018vbk} for the case when non-Gaussianity effects cannot be neglected.} The probability of large fluctuations leading to the formation of PBHs with mass $M$ is then given by:
\be
\beta_{\rm f}(M)=\int_{\zeta_c}^\infty  \frac{1}{\sqrt{2\pi}\,\sigma_\M}\, e^{-\frac{\zeta^2}{2\sigma_\M^2}}\,d\zeta\ ,
\label{eq:beta}
\ee
where $\zeta_c$ denotes the critical value for the collapse into a PBH to take place and plays a fundamental r\^ole in this discussion. It is usually taken to be close to unity, see e.g. \cite{Ballesteros:2017fsr,Motohashi:2017kbs,Germani:2017bcs}.\footnote{We note that some authors \cite{Garcia-Bellido:2017mdw,Ezquiaga:2017fvi} take it to be of the order $10^{-1}$ or $10^{-2}$. Given the exponential dependence of $\beta$ on $\zeta_c$ this significantly decreases the level of tuning required of the inflationary potential in models where PBHs are created within single field inflation.} For such a Gaussian distribution $\sigma_\M^2\sim \langle\zeta \zeta\rangle$ which on CMB scales is $\mc{O}(10^{-9})$. As we will show below, $\sigma_\M \ll \zeta_c$ and so we can approximate (\ref{eq:beta}) as:
\be
\beta_{\rm f}(M)\sim \frac{\sigma_\M}{\sqrt{2 \pi}\,\zeta_c}\,e^{-\frac{\zeta_c^2}{2 \sigma_\M^2}}\,.
\label{beta}
\ee
If PBHs are to be a significant fraction of dark matter, the fluctuations that give rise to them must not be too rare, meaning that $\sigma_\M$ cannot be arbitrarily smaller than $\zeta_c$. This implies that on smaller distance scales the scalar power spectrum must be orders of magnitude larger than on CMB scales. Let us quantify this statement and discuss how it may be achieved in single field models of inflation.

The mass of a PBH forming when a large density perturbation re-enters the horizon is assumed to be proportional to the horizon mass:
\be
M= \gamma \,\frac{4\pi}{3} \frac{\rho_{\rm tot}}{H^3}\Big|_{\rm f}=4\pi\gamma\,\frac{M_p^2}{H_{\rm f}}\ ,
\label{eq:Mpbh}
\ee
where $\gamma$ is a correction factor which depends on the details of the gravitational collapse and $H_{\rm f}$ denotes the Hubble parameter at the moment the perturbation re-enters the horizon. Noting that PBHs behave as matter, the fraction of the total energy density in PBHs at formation time (\ref{betaf}) can be related to the present PBH energy density as:
\be
\beta_{\rm f}(M) = \left(\frac{H_0}{H_{\rm f}}\right)^2 \frac{\Omega_\DM}{a_{\rm f}^3}\,f_\PBH(M)\,,
\label{betaredshift}
\ee
where $a_{\rm f}$ denotes the scale factor at PBH formation time, $H_0$ is the Hubble scale today, $\Omega_\DM =0.26$ is the present fraction of the total energy density in dark matter and $f_\PBH (M)$ is the fraction of the total dark matter energy density in PBHs with mass $M$ today. PBHs in the low mass region, which can be interesting dark matter candidates, get formed before matter-radiation equality in an epoch of radiation dominance. Hence the Hubble scale at PBH formation redshifts as:
\be
H_{\rm f}^2 = \Omega_{\rm r}\,\frac{H_0^2}{a_{\rm f}^4}\left(\frac{g_{*{\rm f}}}{g_{*0}}\right)^{1/3}\,,
\label{Hf}
\ee
where $\Omega_{\rm r}=8\times 10^{-5}$ is the present fraction of the total energy density in radiation, while $g_{*0}$ and $g_{*{\rm f}}$ are respectively the number of relativistic degrees of freedom today and at PBH formation time. Combining (\ref{eq:Mpbh}) with (\ref{Hf}), (\ref{betaredshift}) can be rewritten in terms of present day observables and in units of the solar mass $M_\odot$ as \cite{Carr:2009jm, Inomata:2017okj}:
\be
\beta_{\rm f}(M) \simeq \frac{4}{\sqrt{\gamma}}\times 10^{-9} \left(\frac{g_{*{\rm f}}}{g_{*0}}\right)^{1/4} \sqrt{\frac{M}{M_\odot}}\,f_\PBH (M)\,.
\label{betanew}
\ee
We can now get an estimate of the level of enhancement of the power spectrum required to have PBHs which constitute a significant fraction of dark matter. Setting $\gamma=1$ to be conservative and assuming that only SM degrees of freedom are present so that $g_{*0}=3.36$ and $g_* = 106.75$, if PBHs with mass $M$ constitute all of dark matter, i.e. $f_\PBH(M)=1$ , (\ref{betanew}) reduces to:
\be
\beta_{\rm f}(M) \simeq 10^{-8} \sqrt{\frac{M}{M_\odot}}\,.
\label{betanew2}
\ee
If we now focus on a mass distribution sharply peaked at $M=10^{-15} M_\odot$, we find $\beta_{\rm f}(M)\simeq 3\times 10^{-16}$. Comparing (\ref{betanew2}) with (\ref{beta}) for $\zeta_c=1$, we finally obtain $\sigma_\M=0.12$. This implies that the scalar power spectrum must be enhanced to $\mc{O}(10^{-2})$, a value 7 orders of magnitude larger that its value on CMB scales.\footnote{Had we assumed $\zeta_c=0.1$, we would have found $\sigma=0.012$, in agreement with the estimates of \cite{Garcia-Bellido:2017mdw,Ezquiaga:2017fvi}. This corresponds to an enhancement of the power spectrum by 5 orders of magnitude between PBH and CMB scales and requires less tuning of the inflationary potential.} This large enhancement can in principle be achieved within single field inflationary models by inducing an extremely flat and sufficiently long region in the scalar potential. Therefore the problem of PBHs in single field inflation is one of having a sufficiently rich structure in the scalar potential and the freedom to tune in a flat plateau in the later part of inflation.

Let us finally make two important observations:
\bi
\item In the estimate above of the enhancement of the power spectrum, we considered PBHs with a given mass $M$. However, more generically, the PBH mass function is broadly peaked, and so the fraction of the total dark matter density in PBHs looks like \cite{Carr:2017jsz, Sasaki:2018dmp}:
\be
f_\PBH = \int d f_\PBH(M) = \int \frac{d f_\PBH(M)}{d\ln M} \,d\ln M\,,
\ee
where $d f_\PBH(M)$ is the fraction of PBHs with mass between $M$ and $M+d\ln M$, and the integration domain is bounded below by Hawking evaporation of very light PBHs and above by the mass corresponding to PBHs which re-enter the horizon after matter-radiation equality, see e.g. \cite{Motohashi:2017kbs}. 

\item Assuming that the Hubble scale during inflation $H_{\rm inf}$ is approximately constant, (\ref{eq:Mpbh}) and (\ref{Hf}) can be used to write the number of efoldings between CMB and PBH horizon exit as \cite{Motohashi:2017kbs}:
\bea
\Delta N_\CMB^\PBH&=& \ln \left(\frac{a_\PBH H_{\rm inf}}{a_\CMB H_{\rm inf}}\right) = \ln \left(\frac{a_{\rm f} H_{\rm f}}{0.05\,{\rm Mpc}^{-1}}\right) \nn \\
&=& 18.4-\frac{1}{12}\ln\left(\frac{g_*}{g_{*0}}\right)+\frac12\ln\gamma-\frac12\ln \left(\frac{M}{M_\odot}\right)\,.
\eea
Setting again $\gamma=1$, $g_{*0}=3.36$ and $g_* = 106.75$ as in the SM case, the formation of PBHs with masses in the $[10^{-16},10^{-14}]\, M_\odot$ range implies that PBH scales leave the horizon approximately $34.2$ to $36.5$ efoldings after the CMB scales.
\ei

\section{PBHs from Fibre inflation}
\label{PBHFibre}

\begin{figure}[h]
\begin{center}
\includegraphics[width=0.6\textwidth]{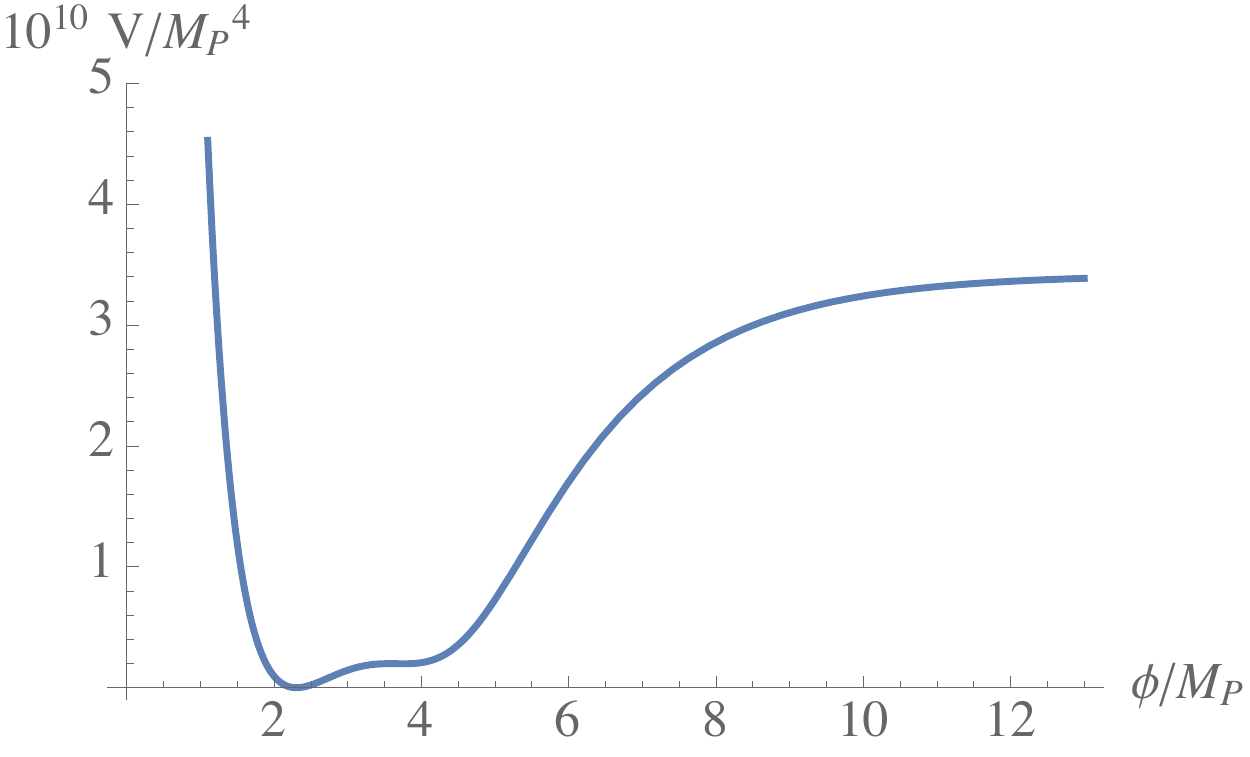}
\caption{Scalar potential for the parameter set $\mathcal{P}_2$ of Tab. \ref{tab:pars}.}
\label{fig:V}
\end{center}
\end{figure}

In order to produce a significant fraction of PBHs from inflationary density perturbations, we shall use the rich structure of the fibre inflation potential (\ref{eq:Vinf}) to induce a near inflection point close to the minimum as depicted in Fig. \ref{fig:V} .

Based on the scaling of each term in eq. \eqref{eq:Vinf} with the fibre modulus $\tau_\K3$ one can see that the second and third terms dominate at small field values and induce a minimum for the modulus around:
\be
\langle\tau_\K3\rangle\sim\frac{C_\W B_\W^2}{(\sqrt{C_\W}-\sqrt{A_\W})^2}\ . 
\ee

The forth term, being proportional to $\tau_\K3$, dominates $V$ at large field values, while the fifth term has a maximum at $\frac{2^{2/3}}{(R_\W / \vo)^{2/3}}$ and scales as $-\tau_\K3$ at small and as $-\tau_\K3^{-1/2}$ at large field values respectively. It is this last term that will be instrumental in generating the enhancement in the scalar power spectrum that will ultimately lead to the formation of primordial black holes in this setup. This can be achieved for certain values of $G_\W$ and $R_\W$ such that the potential has a very flat region close to the post-inflationary minimum as illustrated in Fig. \ref{fig:V}. 

Since in slow-roll $P_k\propto {H^2}/{\epsilon_V}$,  an enhancement of the scalar power spectrum is in principle possible in the limit $\epsilon_V\equiv \frac{V_\phi^2}{2 V^2}\rightarrow 0$. Actually the situation is a little more involved since in the plateau the dynamics of the Universe deviates significantly from  slow-roll, a fact that has been pointed out in \cite{Germani:2017bcs} (see also \cite{Motohashi:2017kbs}), and that calls for a more careful analysis of the observational signatures of such models, see e.g. \cite{Ballesteros:2017fsr}. Observables must therefore be computed from solutions to the Mukhanov-Sasaki equation for the rescaled curvature perturbations: 
\be
u_k''(\eta)+\left(k^2-z''/z\right)u_k(\eta)=0\ ,
\label{eq:MS}
\ee
where $\eta$ denotes conformal time, $z\equiv \sqrt{2 \epsilon} \ a$ from which we find that the effective mass of the curvature perturbations is:
\be
\frac{z''}{z}=(a H)^2\left[2-\epsilon+\frac32 \eta -\frac12 \epsilon \eta+\frac14 \eta^2+\frac12 \eta \kappa\right],
\label{eq:m}
\ee
where:
\be
\epsilon=-\frac{\dot{H}}{H^2} \qquad,\qquad \eta=\frac{\dot{\epsilon}}{\epsilon H} \qquad,\qquad \kappa=\frac{\dot{\eta}}{\eta H}\ ,
\ee
are the Hubble slow-roll parameters.

One assumes that deep inside the horizon, the perturbations behave as if in flat space, which fixes the initial conditions to be of the Bunch-Davies type \cite{Bunch:1978yq}:
\be
\lim_{k\eta \rightarrow -\infty} u_k(\eta)=\frac{e^{-i k \eta}}{\sqrt{2 k}}\ .
\ee
This determines the solution to be given by a Hankel function of the first kind:
\be
u_k(\eta)= \frac{\sqrt{-\pi \eta}}{2}H_{\nu}^{(1)}(-k \eta)\ ,
\ee
with index $\nu$ determined from eq. \eqref{eq:m} once a given background is chosen.

\begin{figure}[h]
\begin{center}\centering
	\begin{minipage}[b]{0.49\linewidth}
	\centering
	\includegraphics[width=1\textwidth]{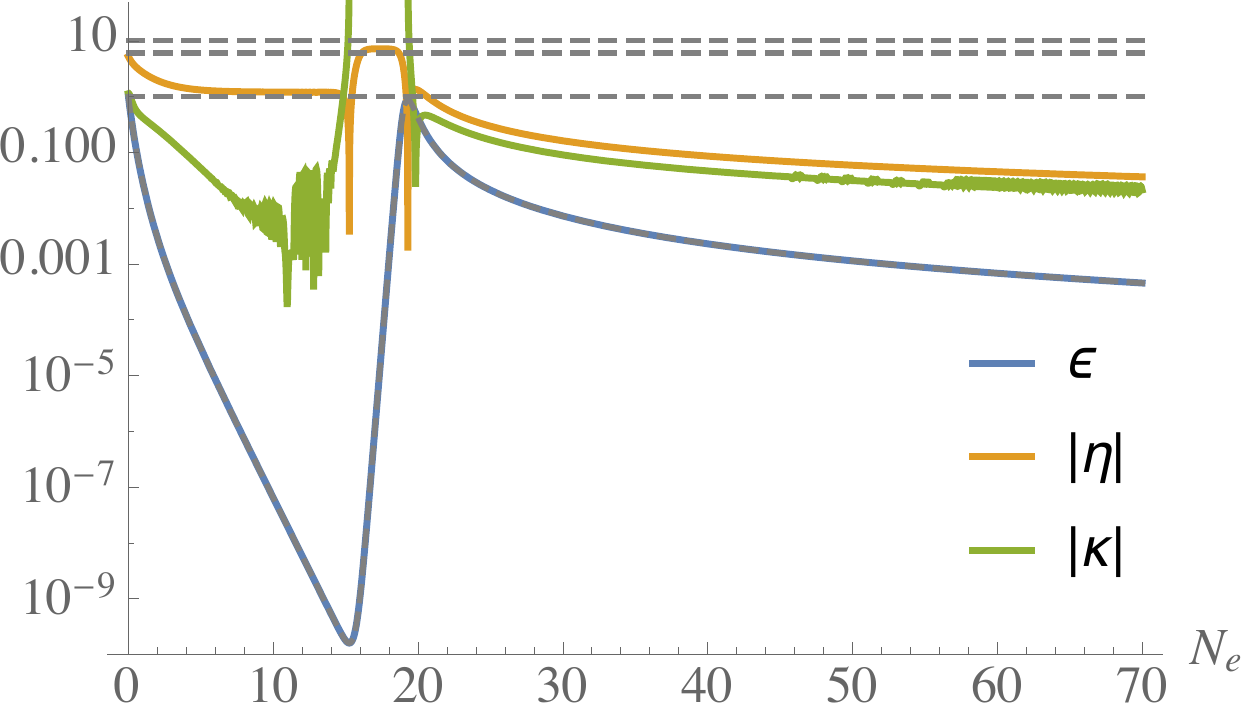}
    \end{minipage}
	\hspace{0.05cm}
	\begin{minipage}[b]{0.49\linewidth}
	\centering
	\includegraphics[width=1\textwidth]{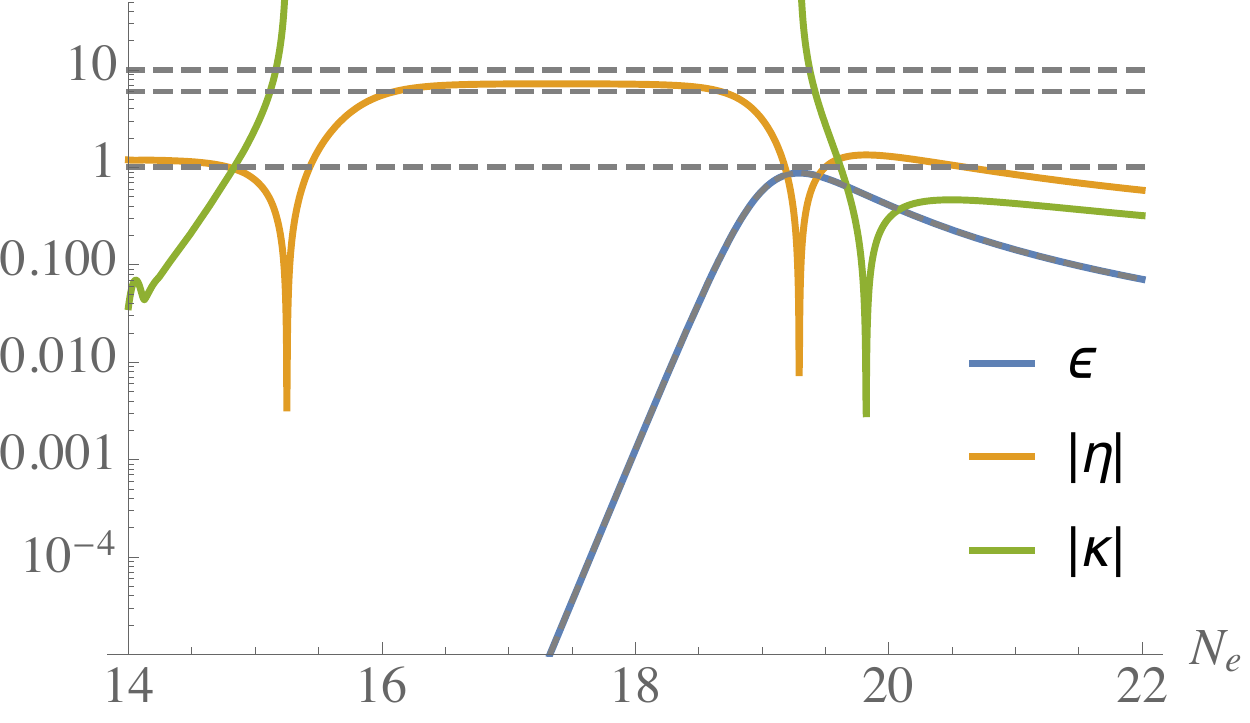}
	\end{minipage}
	\hspace{0.05cm}
\centering
\caption{Slow-roll parameters as functions of the number of efoldings $N$ from the end of inflation for parameter set $\mathcal{P}_2$. It is clear that the background evolves from slow-roll ($N_e>19$) to constant-roll ($15<N_e<19$) and back to slow-roll again ($N_e<15$). Dashed lines represent $10$, $6$ and $1$. }
\label{fig:srpars}
\end{center}
\end{figure}

For comparison with observations one is interested in the dimensionless power spectrum, defined as: 
\be
P_k=\frac{k^3}{2 \pi^2 }\left |\frac{u_k}{z}\right|^2\ ,
\label{eq:Pk}
\ee
which in the superhorizon limit $k \eta \rightarrow 0$ can be written as: 
\be
P_k=\frac{H^2}{8 \pi^2 \epsilon}\frac{2^{2\nu-1}|\Gamma(\nu)|^2}{\pi}\left(\frac{k}{a H}\right)^{3-2\nu}\ .
\label{eq:Pk2}
\ee
On CMB scales this is bound to be $P_k\big |_\CMB= 2\times10^{-9}$ and as shown in Sec. \ref{sec:PBH} it must be significantly enhanced on smaller scales if PBHs are to be significant fraction of all dark matter.\\
Up to this point the discussion of the behaviour of the perturbations assumed nothing about the type of background in which they evolve.  In order to produce a significant amount of PBH from a inflection point in single field inflation,  we will see that the universe has to evolve from a slow-roll inflation phase into a transient constant-roll background, where the scalar field acceleration plays an important role. These backgrounds are characterised by the parameter $\alpha$ defined as \cite{Martin:2012pe,  Motohashi:2014ppa}:
\be
\ddot{\phi} \equiv-(3+\alpha)H \dot{\phi} \ .
\ee
Solutions with $\alpha=0$ are called ultra slow-roll \cite{Kinney:2005vj,Tsamis:2003px,Namjoo:2012aa}, whereas vanilla slow-roll inflation corresponds to $\alpha=-3$. The transient constant-roll period arises due the presence of an extremely flat region in the potential that causes the scalar field to brake upon reaching it, leading to a non negligible acceleration in the Klein-Gordon equation and consequently a departure from the slow-roll background. This behaviour is illustrated in Fig. \ref{fig:srpars} where we plot the evolution of the slow-roll parameters for evolution in the potential of Fig. \ref{fig:V}, corresponding to the parameter set $\mathcal{P}_2$ of Tab. \ref{tab:pars}. It is evident that the system undergoes a transition from slow-roll ($N_e>19$) to constant-roll ($15<N_e<19$) and finally to a large $\eta$ slow-roll phase ($N_e<15$).

In slow-roll $\epsilon, \eta, \kappa \ll1$, and consequently the effective mass takes the form $z''/z\approx \frac{2}{\eta^2}\left( 2+3\epsilon+3\eta \right)$, or equivalently $\nu=3/2+\epsilon +\eta/2$. One can then see that the curvature perturbations $\zeta=u/z$ remain constant on super-horizon scales and the two point function can therefore be evaluated at horizon crossing, yielding the familiar slow-roll result:
\be
P_k=\frac{H^2}{8 \pi^2 \epsilon}\Big |_{k=a H}\ .
\label{eq:Pk_SR}
\ee
Eq. \eqref{eq:Pk2} also captures the momentum dependence of the two point function, which can be written in terms of the spectral index $n_s$ and its running $\alpha$, given by:
\be
n_s\equiv \frac{d \ln P_k}{d \ln k}=1-2\epsilon-\eta\,,
\ee
and:
\be
\frac{d n_s}{ d\ln k}= -2 \epsilon \eta - \eta \kappa\ .
\ee
Both these quantities are subject of tight observational constraints \cite{Ade:2015lrj}. For this work we take:
\be
n_s= 0.9650\pm 0.0050\qquad\text{and}\qquad \frac{d n_s}{ d\ln k}=-0.009 \pm 0.008
\ee
at $68\%$CL and at a scale $k_*=0.05\  \text{Mpc}^{-1}$.

In the transient constant-roll regime one has $\eta\approx-2(3+\alpha-\epsilon)$ which implies $\epsilon\propto a^{-2(3+\alpha)}$. In the cases we consider $\alpha\in[0,1]$. In such a background the super-horizon behaviour of the power spectrum is determined by:
\be
P_k\propto H^ {|2\alpha+3|-1} a^ {3+2\alpha+|3+2\alpha|} \ .
\ee
Note that since $\epsilon$ is small and decreasing rapidly with the expansion (for $\alpha>-3$), one can take $H$ to be constant. We therefore see that for $-3\le\alpha< -3/2$ the curvature perturbations are frozen beyond the horizon (this includes the previously discussed case of slow-roll inflation, $\alpha=-3$), whereas for $\alpha> -3/2$, $P_k\propto a^ {2(3+2\alpha)}$, signaling the presence of a growing solution to the MS equation, and the breakdown of the approximation of Eq. \eqref{eq:Pk_SR}. In order to determine the two-point function in such backgrounds one must therefore solve the MS equation and evaluate $P_k$ at the end of inflation.

\begin{figure}[h]
\begin{center}
\includegraphics[width=0.7\textwidth]{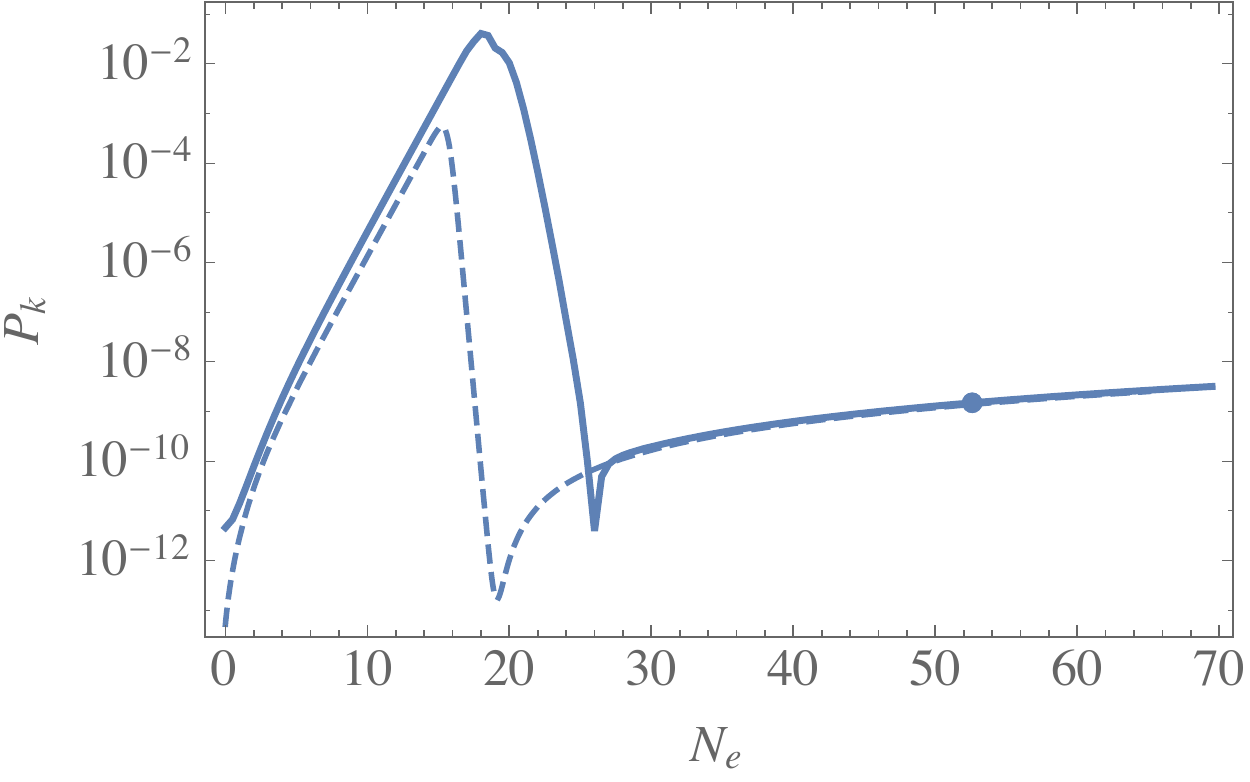}
\caption{Power spectrum Eq. \eqref{eq:Pk} for the potential of Fig. \ref{fig:V} with parameter set $\mathcal{P}_2$. The dashed line represents the slow-roll estimate of Eq. \eqref{eq:Pk_SR} while the continuous line is obtained from the solutions to the MS equation. The circle correspond to the CMB scales if the peak is to be associated with PBH of mass $M=10^{-14} M_\odot$.}
\label{fig:Pk}
\end{center}
\end{figure}

\begin{figure}[h]
\begin{center}\centering
	\begin{minipage}[b]{0.49\linewidth}
	\centering
	\includegraphics[width=0.99\textwidth]{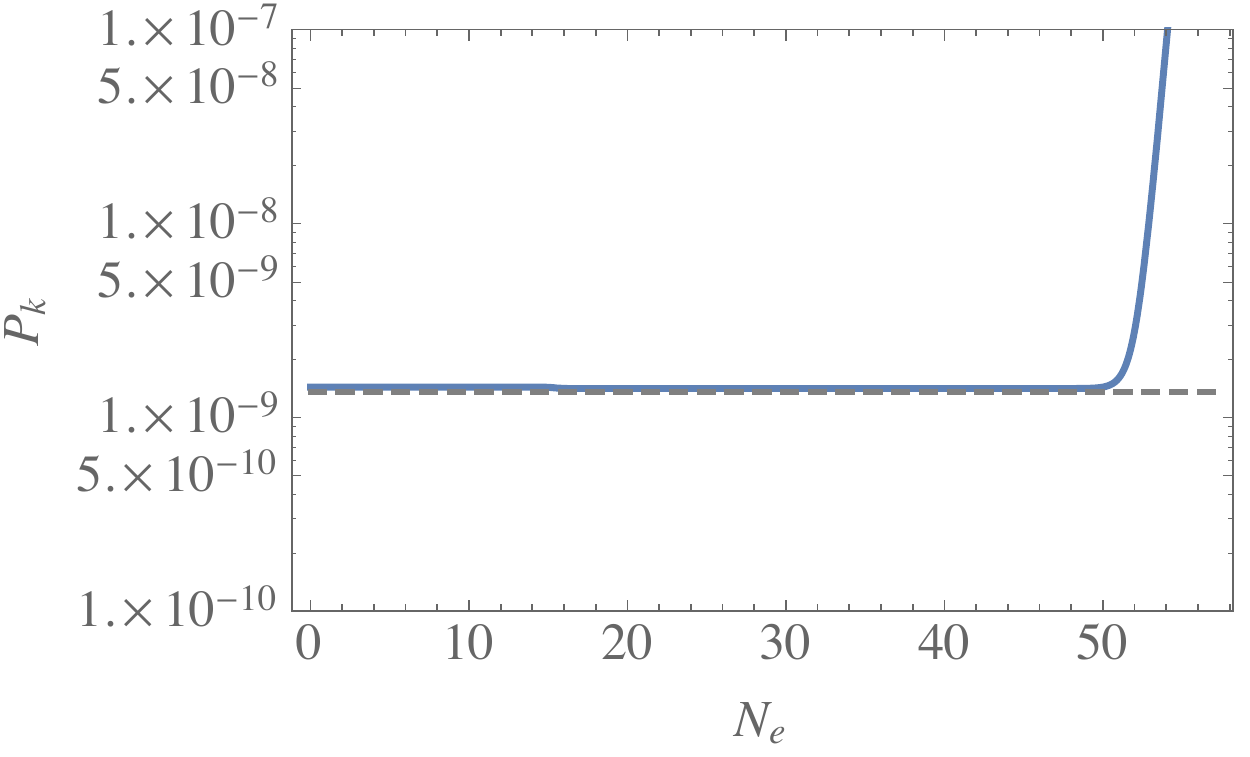}
    \end{minipage}
	\hspace{0.05cm}
	\begin{minipage}[b]{0.49\linewidth}
	\centering
	\includegraphics[width=0.92\textwidth]{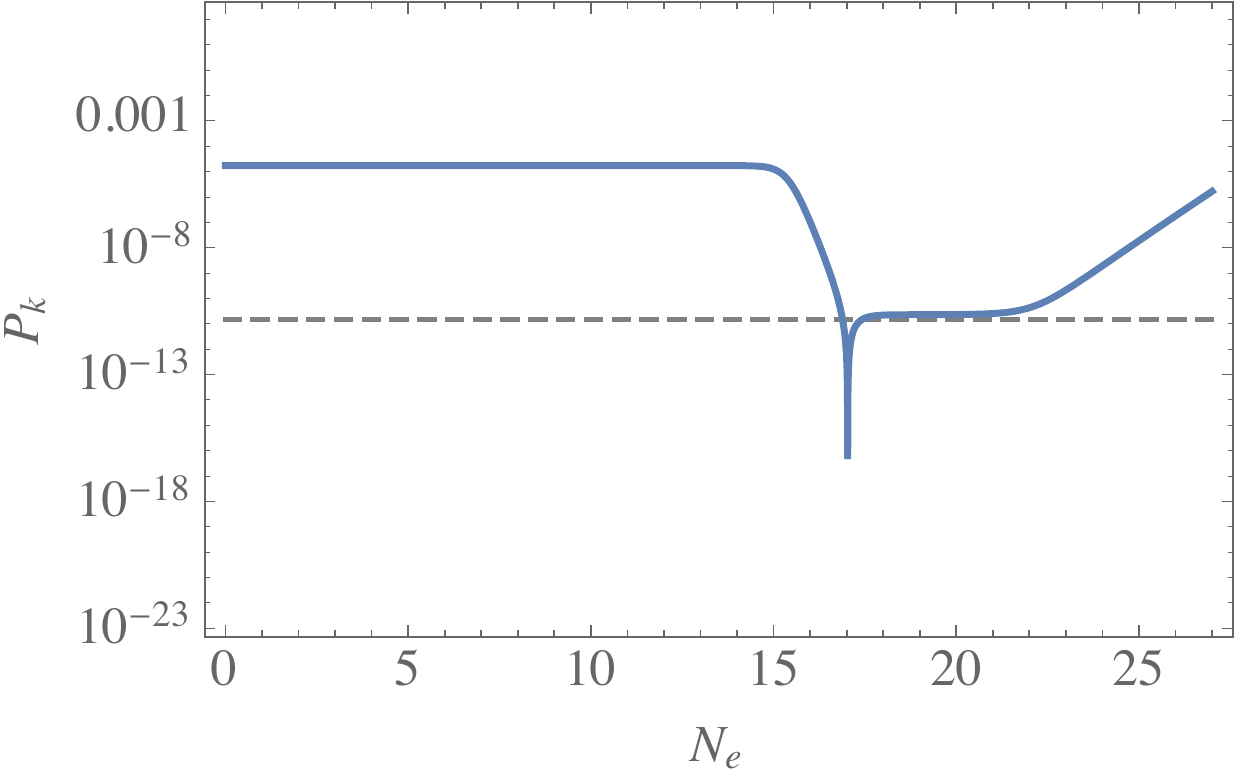}
	\end{minipage}
	\hspace{0.05cm}
\centering
\caption{Evolution of the curvature perturbations on different scales for example $\mathcal{P}_2$. On the left panel a mode that corresponds to CMB scales, leaving the horizon $53$ efoldings before the end of inflation, and keeping a constant value thereafter. On the right panel a mode that corresponds to PBH scales, leaving the horizon 22 efoldings before the end of inflation and undergoing super-horizon growth during the constant-roll phase. In both plots the dashed line corresponds to the slow-roll estimate of eq. \eqref{eq:Pk_SR}.}
\label{fig:Pk_SR_v_USR}

\end{center}
\end{figure}

In Fig. \ref{fig:Pk} we plot the power spectrum for scalar perturbations for the potential of Fig. \ref{fig:V} (parameter set $\mc{P}_2$) calculated from the solutions of Eq. \eqref{eq:MS} (continuous line) and the slow-roll estimate of Eq. \eqref{eq:Pk_SR} (dashed line). As expected the slow-roll approximation breaks down for modes that cross the horizon close to the onset of the constant-roll phase. Crucially for the production of PBH, the slow-roll result underestimates the power spectrum by several orders of magnitude in this range of momentum modes. This is to be expected given the existence of a growing mode solution in constant-roll backgrounds with $\alpha\in[0,1]$. In Fig. \ref{fig:Pk_SR_v_USR} we plot the evolution of the power spectrum for modes leaving the horizon $53$ and $22$ efoldings before the end of inflation, corresponding to CMB and PBH scales respectively. While both scales are affected by the growing mode during the constant roll phase, their superhorizon growth is determined by $\frac{k}{aH}$ at the onset of the constant roll period. This quantity is minute for modes on CMB scales but not for those on the PBH region. As a result the CMB mode essentially follow the slow-roll estimate of Eq. \eqref{eq:Pk_SR} after crossing the horizon,  while on small scales we see that there is a large amplification of $P_k$  leading to a breakdown of the slow-roll approximation. Finally let us note that the $\mc{O}(1)$ deviation from the slow-roll estimate for modes on the smallest scales ($N_e\lesssim15$) can be attributed to the fact that in the final phase of expansion before the end of inflation, $\eta=\mc{O}(1)$.

In Tables \ref{tab:pars} and \ref{tab:obs} we present three numerical examples corresponding to cases where  all DM is composed of $10^{-14} M_\odot$ PBHs, assuming $\zeta_c=1$. We stress that the choices of the microphysical parameters are in line with expectations (including the small values of $G_W$ and $R_W$) and that the desire to have PBH DM does not constrain the compactification volume, which varies by several orders of magnitude in between the three examples. All examples lead to a spectral index that is $2$ to $3$ sigma redder than the current best fit, while giving rise to a spectral index running and a tensor fraction that are in line with current bounds.

\begin{table}[htp]
\begin{center}
\begin{tabular}{c||c|c|c|c|c||c|c}
&$C_W$&$A_W$&$B_W$&$G_W/\langle\mathcal{V} \rangle$&$R_W/\langle\mathcal{V} \rangle$&$\langle \tau_{K3}\rangle $&$\langle \mathcal{V}\rangle $\\
\hline
$\mathcal{P}_1$& $1/10$ & $2/100$ & $1$ & $1.303386\times10^{-3}$ & $6.58724\times10^{-3}$& $3.89$&$107.3$\\
$\mathcal{P}_2$& $4/100$& $2/100$& $1$& $3.080548\times 10^{-5}$ &$7.071067 \times 10^{-4}$ &$14.30 $&$1000$ \\
$\mathcal{P}_3$& $1.978/100$&$1.65/100$&$1.01$ &$9.257715\times 10^{-8}$ &$1.414\times10^{-5}$ &$168.03$ &$5\times10^4$ \\
\end{tabular}
\end{center}
\caption{Examples of parameters leading to the production of PBHs with a mass peaked at  $10^{-14} M_\odot$, together with geometrical compactification data. All examples exhibit $D_W=0$. Note that the small values of $G_W$ and $R_W$ are in line with their microscopic origin as explained in Sec. \ref{FIReview}.}
\label{tab:pars}
\end{table}

\begin{table}[htp]
\begin{center}
\begin{tabular}{c||c|c|c||c|c}
&$n_s$ & $r$ & $\frac{d n_s}{d \log k}$ &  $\Delta N_{CMB}^{PHB}$ & $P_k|_{\rm peak}$ \\
\hline
$\mathcal{P}_1$&0.9457&$0.015$ &$-0.0017$&$34.5$&$0.01365$ \\
$\mathcal{P}_2$&0.9437&$0.015$ &$-0.0017$&$34.5$&$0.03998$ \\
$\mathcal{P}_3$&0.9457&$0.015$ &$-0.0019$&$34.5$&$0.013341$ \\
\end{tabular}
\end{center}
\caption{Inflationary observables on CMB and PBH scales for the examples of table \ref{tab:pars}.}
\label{tab:obs}
\end{table}

\section{Conclusions}
\label{Concl}

In this paper we have presented the first explicit example of a string inflationary model which is consistent with cosmological observations at CMB scales and, at the same time, can generate PBHs at small distance scales via an efficient enhancement of the power spectrum due to a period of ultra slow-roll. Our model leads to PBHs in the low-mass region where they constitute a significant fraction of the total dark matter abundance. 

Three interesting features of fibre inflation models relevant for PBH formation are the following: ($i$) the coefficients of the different contributions to the inflationary potential depend on microscopic parameters like background fluxes and Calabi-Yau intersection numbers which take different values in the string landscape, and so give a very large tuning freedom that can be used to generate a near inflection point; ($ii$) the potential enjoys an approximate Abelian rescaling symmetry inherited from the underlying extended no-scale structure which suppresses quantum corrections to the inflationary dynamics; ($iii$) the contribution to the inflationary potential responsible for the emergence of a near inflection point at large momentum scales has been derived in global embeddings of fibre inflation models in explicit Calabi-Yau compactifications with chiral brane setup compatible with moduli stabilisation. 

Moreover our model is characterised by a trans-Planckian field range during inflation, and so it predicts a large tensor-to-scalar ratio of order $r\sim 0.01$ which might be detected by the next generation of cosmological measurements. Similarly to previous works on PBH formation in single-field inflation \cite{Ezquiaga:2017fvi, Ballesteros:2017fsr}, the scalar spectral index turns out to be a bit too red since it is more than $3\sigma$ away from the Planck reference value. This tension might be resolved by the inclusion of non-zero neutrino masses which might make our result compatible with CMB data within just $2\sigma$ \cite{Gerbino:2016sgw}. The tension in the values of $n_s$ can also decrease in compactifications where the approach to the inflationary plateau occurs faster than the $1/\sqrt{\tau_{K3}}$ considered here, a possibility in potentials of the form of Eq. \eqref{PotFI}. Another interesting cosmological observable in our model is the running of the spectral index which turns out to be sizable.

In this paper we investigated the possibility to generate PBHs from string inflation by taking the most conservative point of view since we focused on models which are effectively single-field and, above all, we considered PBH formation with horizon re-entry in a radiation dominated era. However, two generic features of string compactifications are the presence of several scalar fields which might play an important r\^ole during inflation \cite{Dimopoulos:2005ac, Easther:2005zr, Bond:2006nc, Berglund:2009uf, BlancoPillado:2009nw, Burgess:2010bz, Cicoli:2012cy, Cicoli:2014sva}, and light moduli with only gravitational couplings to ordinary matter which are long-lived and tend to give rise to early periods of matter domination \cite{Acharya:2008bk, Kane:2011ih, Cicoli:2012aq, Higaki:2012ar, Allahverdi:2013noa, Dutta:2014tya, Cicoli:2016olq, Allahverdi:2016yws}. Hence in the future it would be very interesting to study the impact on PBH formation in string models of additional light fields, like for example the axionic partner of the inflaton of fibre inflation models.

We finally mention the fact that PBHs can be generated with the required efficiency only if $\delta\rho \sim 0.1\, \rho$ at small distance scales. It would therefore be important to perform a more careful analysis of stochastic effects since the perturbative expansion might not be fully justified \cite{Pattison:2017mbe}. Finally we stress that non-gaussianities in large momentum fluctuations might alter significantly the PBH production mechanism and, in turn, their present abundance \cite{Franciolini:2018vbk}. We leave a deeper study of both stochastic and non-gaussianity effects for future investigation.

\section*{Acknowledgements}

We are thankful to R. Allahverdi, B. Dutta and C.~Germani for helpful discussions. VAD thanks Matteo Capozi for useful programming related discussions and the Max-Planck-Institut f\"{u}r Physik for hospitality and support in the  final stages of this work.

\bibliographystyle{JHEP}
\bibliography{References}

\providecommand{\href}[2]{#2}\begingroup\raggedright\begin{thebibliography}{10}

\bibitem{Sasaki:2018dmp}
  M.~Sasaki, T.~Suyama, T.~Tanaka and S.~Yokoyama,
  Class.\ Quant.\ Grav.\  {\bf 35} (2018) no.6,  063001
  doi:10.1088/1361-6382/aaa7b4
  [arXiv:1801.05235 [astro-ph.CO]].


\bibitem{Hawking:1974rv}
  S.~W.~Hawking,
  Nature {\bf 248} (1974) 30.
  doi:10.1038/248030a0


\bibitem{Ali-Haimoud:2016mbv}
  Y.~Ali-Haïmoud and M.~Kamionkowski,
  Phys.\ Rev.\ D {\bf 95} (2017) no.4,  043534
  doi:10.1103/PhysRevD.95.043534
  [arXiv:1612.05644 [astro-ph.CO]].


\bibitem{Poulin:2017bwe}
  V.~Poulin, P.~D.~Serpico, F.~Calore, S.~Clesse and K.~Kohri,
  Phys.\ Rev.\ D {\bf 96} (2017) no.8,  083524
  doi:10.1103/PhysRevD.96.083524
  [arXiv:1707.04206 [astro-ph.CO]].


\bibitem{Gaggero:2016dpq}
  D.~Gaggero, G.~Bertone, F.~Calore, R.~M.~T.~Connors, M.~Lovell, S.~Markoff and E.~Storm,
  Phys.\ Rev.\ Lett.\  {\bf 118} (2017) no.24,  241101
  doi:10.1103/PhysRevLett.118.241101
  [arXiv:1612.00457 [astro-ph.HE]].


\bibitem{Hawking:1971ei}
  S.~Hawking,
  Mon.\ Not.\ Roy.\ Astron.\ Soc.\  {\bf 152} (1971) 75.


\bibitem{Carr:1974nx}
  B.~J.~Carr and S.~W.~Hawking,
  Mon.\ Not.\ Roy.\ Astron.\ Soc.\  {\bf 168} (1974) 399.


\bibitem{Ade:2015lrj}
  P.~A.~R.~Ade {\it et al.} [Planck Collaboration],
  Astron.\ Astrophys.\  {\bf 594} (2016) A20
  doi:10.1051/0004-6361/201525898
  [arXiv:1502.02114 [astro-ph.CO]].


\bibitem{Kawasaki:2012wr}
  M.~Kawasaki, N.~Kitajima and T.~T.~Yanagida,
  Phys.\ Rev.\ D {\bf 87} (2013) no.6,  063519
  doi:10.1103/PhysRevD.87.063519
  [arXiv:1207.2550 [hep-ph]].


\bibitem{Carr:2016drx}
  B.~Carr, F.~Kuhnel and M.~Sandstad,
  Phys.\ Rev.\ D {\bf 94} (2016) no.8,  083504
  doi:10.1103/PhysRevD.94.083504
  [arXiv:1607.06077 [astro-ph.CO]].


\bibitem{Garcia-Bellido:2016dkw}
  J.~Garcia-Bellido, M.~Peloso and C.~Unal,
  JCAP {\bf 1612} (2016) no.12,  031
  doi:10.1088/1475-7516/2016/12/031
  [arXiv:1610.03763 [astro-ph.CO]].


\bibitem{Ivanov:1994pa}
  P.~Ivanov, P.~Naselsky and I.~Novikov,
  Phys.\ Rev.\ D {\bf 50} (1994) 7173.
  doi:10.1103/PhysRevD.50.7173


\bibitem{Garcia-Bellido:2017mdw}
  J.~Garcia-Bellido and E.~Ruiz Morales,
  Phys.\ Dark Univ.\  {\bf 18} (2017) 47
  doi:10.1016/j.dark.2017.09.007
  [arXiv:1702.03901 [astro-ph.CO]].


\bibitem{Ezquiaga:2017fvi}
  J.~M.~Ezquiaga, J.~Garcia-Bellido and E.~Ruiz Morales,
  Phys.\ Lett.\ B {\bf 776} (2018) 345
  doi:10.1016/j.physletb.2017.11.039
  [arXiv:1705.04861 [astro-ph.CO]].


\bibitem{Germani:2017bcs}
  C.~Germani and T.~Prokopec,
  Phys.\ Dark Univ.\  {\bf 18} (2017) 6
  doi:10.1016/j.dark.2017.09.001
  [arXiv:1706.04226 [astro-ph.CO]].


\bibitem{Ballesteros:2017fsr}
  G.~Ballesteros and M.~Taoso,
  Phys.\ Rev.\ D {\bf 97} (2018) no.2,  023501
  doi:10.1103/PhysRevD.97.023501
  [arXiv:1709.05565 [hep-ph]].


\bibitem{Hertzberg:2017dkh}
  M.~P.~Hertzberg and M.~Yamada,
  arXiv:1712.09750 [astro-ph.CO].


\bibitem{Baumann:2009ni}
  D.~Baumann and L.~McAllister,
  Ann.\ Rev.\ Nucl.\ Part.\ Sci.\  {\bf 59} (2009) 67
  doi:10.1146/annurev.nucl.010909.083524
  [arXiv:0901.0265 [hep-th]].


\bibitem{Burgess:2013sla}
  C.~P.~Burgess, M.~Cicoli and F.~Quevedo,
  JCAP {\bf 1311} (2013) 003
  doi:10.1088/1475-7516/2013/11/003
  [arXiv:1306.3512 [hep-th]].


\bibitem{Pajer:2013fsa}
  E.~Pajer and M.~Peloso,
  Class.\ Quant.\ Grav.\  {\bf 30} (2013) 214002
  doi:10.1088/0264-9381/30/21/214002
  [arXiv:1305.3557 [hep-th]].


\bibitem{Cicoli:2011zz}
  M.~Cicoli and F.~Quevedo,
  Class.\ Quant.\ Grav.\  {\bf 28} (2011) 204001
  doi:10.1088/0264-9381/28/20/204001
  [arXiv:1108.2659 [hep-th]].


\bibitem{Burgess:2014tja}
  C.~P.~Burgess, M.~Cicoli, F.~Quevedo and M.~Williams,
  JCAP {\bf 1411} (2014) 045
  doi:10.1088/1475-7516/2014/11/045
  [arXiv:1404.6236 [hep-th]].


\bibitem{Burgess:2016owb}
  C.~P.~Burgess, M.~Cicoli, S.~de Alwis and F.~Quevedo,
  JCAP {\bf 1605} (2016) no.05,  032
  doi:10.1088/1475-7516/2016/05/032
  [arXiv:1603.06789 [hep-th]].


\bibitem{Kachru:2003aw}
  S.~Kachru, R.~Kallosh, A.~D.~Linde and S.~P.~Trivedi,
  Phys.\ Rev.\ D {\bf 68} (2003) 046005
  doi:10.1103/PhysRevD.68.046005
  [hep-th/0301240].


\bibitem{Kallosh:2014wsa}
  R.~Kallosh and T.~Wrase,
  JHEP {\bf 1412} (2014) 117
  doi:10.1007/JHEP12(2014)117
  [arXiv:1411.1121 [hep-th]].


\bibitem{Bergshoeff:2015jxa}
  E.~A.~Bergshoeff, K.~Dasgupta, R.~Kallosh, A.~Van Proeyen and T.~Wrase,
  JHEP {\bf 1505} (2015) 058
  doi:10.1007/JHEP05(2015)058
  [arXiv:1502.07627 [hep-th]].


\bibitem{Burgess:2003ic}
  C.~P.~Burgess, R.~Kallosh and F.~Quevedo,
  JHEP {\bf 0310} (2003) 056
  doi:10.1088/1126-6708/2003/10/056
  [hep-th/0309187].


\bibitem{Cicoli:2013cha}
  M.~Cicoli, D.~Klevers, S.~Krippendorf, C.~Mayrhofer, F.~Quevedo and R.~Valandro,
  JHEP {\bf 1405} (2014) 001
  doi:10.1007/JHEP05(2014)001
  [arXiv:1312.0014 [hep-th]].


\bibitem{Cicoli:2015ylx}
  M.~Cicoli, F.~Quevedo and R.~Valandro,
  JHEP {\bf 1603} (2016) 141
  doi:10.1007/JHEP03(2016)141
  [arXiv:1512.04558 [hep-th]].


\bibitem{Cicoli:2012fh}
  M.~Cicoli, A.~Maharana, F.~Quevedo and C.~P.~Burgess,
  JHEP {\bf 1206} (2012) 011
  doi:10.1007/JHEP06(2012)011
  [arXiv:1203.1750 [hep-th]].


\bibitem{Cicoli:2008gp}
  M.~Cicoli, C.~P.~Burgess and F.~Quevedo,
  JCAP {\bf 0903} (2009) 013
  doi:10.1088/1475-7516/2009/03/013
  [arXiv:0808.0691 [hep-th]].


\bibitem{Broy:2015zba}
  B.~J.~Broy, D.~Ciupke, F.~G.~Pedro and A.~Westphal,
  JCAP {\bf 1601} (2016) 001
  doi:10.1088/1475-7516/2016/01/001
  [arXiv:1509.00024 [hep-th]].


\bibitem{Cicoli:2016chb}
  M.~Cicoli, D.~Ciupke, S.~de Alwis and F.~Muia,
  JHEP {\bf 1609} (2016) 026
  doi:10.1007/JHEP09(2016)026
  [arXiv:1607.01395 [hep-th]].


\bibitem{Cicoli:2010ha}
  M.~Cicoli and A.~Mazumdar,
  JCAP {\bf 1009} (2010) no.09,  025
  doi:10.1088/1475-7516/2010/09/025
  [arXiv:1005.5076 [hep-th]].


\bibitem{Cabella:2017zsa}
  P.~Cabella, A.~Di Marco and G.~Pradisi,
  Phys.\ Rev.\ D {\bf 95} (2017) no.12,  123528
  doi:10.1103/PhysRevD.95.123528
  [arXiv:1704.03209 [astro-ph.CO]].


\bibitem{Antusch:2017flz}
  S.~Antusch, F.~Cefala, S.~Krippendorf, F.~Muia, S.~Orani and F.~Quevedo,
  JHEP {\bf 1801} (2018) 083
  doi:10.1007/JHEP01(2018)083
  [arXiv:1708.08922 [hep-th]].


\bibitem{Starobinsky:1980te}
  A.~A.~Starobinsky,
  Phys.\ Lett.\  {\bf 91B} (1980) 99.
  doi:10.1016/0370-2693(80)90670-X


\bibitem{Kallosh:2013maa}
  R.~Kallosh and A.~Linde,
  JCAP {\bf 1310} (2013) 033
  doi:10.1088/1475-7516/2013/10/033
  [arXiv:1307.7938 [hep-th]].


\bibitem{Kallosh:2015zsa}
  R.~Kallosh and A.~Linde,
  Comptes Rendus Physique {\bf 16} (2015) 914
  doi:10.1016/j.crhy.2015.07.004
  [arXiv:1503.06785 [hep-th]].


\bibitem{Carrasco:2015pla}
  J.~J.~M.~Carrasco, R.~Kallosh and A.~Linde,
  JHEP {\bf 1510} (2015) 147
  doi:10.1007/JHEP10(2015)147
  [arXiv:1506.01708 [hep-th]].


\bibitem{Cicoli:2016xae}
  M.~Cicoli, F.~Muia and P.~Shukla,
  JHEP {\bf 1611} (2016) 182
  doi:10.1007/JHEP11(2016)182
  [arXiv:1611.04612 [hep-th]].


\bibitem{Cicoli:2017axo}
  M.~Cicoli, D.~Ciupke, V.~A.~Diaz, V.~Guidetti, F.~Muia and P.~Shukla,
  JHEP {\bf 1711} (2017) 207
  doi:10.1007/JHEP11(2017)207
  [arXiv:1709.01518 [hep-th]].


\bibitem{Balasubramanian:2005zx}
  V.~Balasubramanian, P.~Berglund, J.~P.~Conlon and F.~Quevedo,
  JHEP {\bf 0503} (2005) 007
  doi:10.1088/1126-6708/2005/03/007
  [hep-th/0502058].


\bibitem{Cicoli:2008va}
  M.~Cicoli, J.~P.~Conlon and F.~Quevedo,
  JHEP {\bf 0810} (2008) 105
  doi:10.1088/1126-6708/2008/10/105
  [arXiv:0805.1029 [hep-th]].


\bibitem{Kannike:2017bxn}
  K.~Kannike, L.~Marzola, M.~Raidal and H.~Veermäe,
  JCAP {\bf 1709} (2017) no.09,  020
  doi:10.1088/1475-7516/2017/09/020
  [arXiv:1705.06225 [astro-ph.CO]].


\bibitem{Motohashi:2017kbs}
  H.~Motohashi and W.~Hu,
  Phys.\ Rev.\ D {\bf 96} (2017) no.6,  063503
  doi:10.1103/PhysRevD.96.063503
  [arXiv:1706.06784 [astro-ph.CO]].


\bibitem{Sasaki:1986hm}
  M.~Sasaki,
  Prog.\ Theor.\ Phys.\  {\bf 76} (1986) 1036.
  doi:10.1143/PTP.76.1036


\bibitem{Mukhanov:1988jd}
  V.~F.~Mukhanov,
  Sov.\ Phys.\ JETP {\bf 67} (1988) 1297
   [Zh.\ Eksp.\ Teor.\ Fiz.\  {\bf 94N7} (1988) 1].


\bibitem{Becker:2002nn}
  K.~Becker, M.~Becker, M.~Haack and J.~Louis,
  JHEP {\bf 0206} (2002) 060
  doi:10.1088/1126-6708/2002/06/060
  [hep-th/0204254].


\bibitem{Minasian:2015bxa}
  R.~Minasian, T.~G.~Pugh and R.~Savelli,
  JHEP {\bf 1510} (2015) 050
  doi:10.1007/JHEP10(2015)050
  [arXiv:1506.06756 [hep-th]].


\bibitem{Bonetti:2016dqh}
  F.~Bonetti and M.~Weissenbacher,
  JHEP {\bf 1701} (2017) 003
  doi:10.1007/JHEP01(2017)003
  [arXiv:1608.01300 [hep-th]].


\bibitem{Berg:2005ja}
  M.~Berg, M.~Haack and B.~Kors,
  JHEP {\bf 0511} (2005) 030
  doi:10.1088/1126-6708/2005/11/030
  [hep-th/0508043].


\bibitem{Berg:2007wt}
  M.~Berg, M.~Haack and E.~Pajer,
  JHEP {\bf 0709} (2007) 031
  doi:10.1088/1126-6708/2007/09/031
  [arXiv:0704.0737 [hep-th]].


\bibitem{Berg:2014ama}
  M.~Berg, M.~Haack, J.~U.~Kang and S.~Sjörs,
  JHEP {\bf 1412} (2014) 077
  doi:10.1007/JHEP12(2014)077
  [arXiv:1407.0027 [hep-th]].


\bibitem{Haack:2015pbv}
  M.~Haack and J.~U.~Kang,
  JHEP {\bf 1602} (2016) 160
  doi:10.1007/JHEP02(2016)160
  [arXiv:1511.03957 [hep-th]].


\bibitem{Cicoli:2007xp}
  M.~Cicoli, J.~P.~Conlon and F.~Quevedo,
  JHEP {\bf 0801} (2008) 052
  doi:10.1088/1126-6708/2008/01/052
  [arXiv:0708.1873 [hep-th]].


\bibitem{Ciupke:2015msa}
  D.~Ciupke, J.~Louis and A.~Westphal,
  JHEP {\bf 1510} (2015) 094
  doi:10.1007/JHEP10(2015)094
  [arXiv:1505.03092 [hep-th]].


\bibitem{Grimm:2017okk}
  T.~W.~Grimm, K.~Mayer and M.~Weissenbacher,
  JHEP {\bf 1802} (2018) 127
  doi:10.1007/JHEP02(2018)127
  [arXiv:1702.08404 [hep-th]].


\bibitem{Cicoli:2013swa}
  M.~Cicoli, J.~P.~Conlon, A.~Maharana and F.~Quevedo,
  JHEP {\bf 1401} (2014) 027
  doi:10.1007/JHEP01(2014)027
  [arXiv:1310.6694 [hep-th]].


\bibitem{Cicoli:2018tcq}
  M.~Cicoli, D.~Ciupke, C.~Mayrhofer and P.~Shukla,
  arXiv:1801.05434 [hep-th].


\bibitem{Cicoli:2013oba}
  M.~Cicoli, S.~Downes and B.~Dutta,
  JCAP {\bf 1312} (2013) 007
  doi:10.1088/1475-7516/2013/12/007
  [arXiv:1309.3412 [hep-th]].


\bibitem{Pedro:2013pba}
  F.~G.~Pedro and A.~Westphal,
  JHEP {\bf 1404} (2014) 034
  doi:10.1007/JHEP04(2014)034
  [arXiv:1309.3413 [hep-th]].


\bibitem{Cicoli:2014bja}
  M.~Cicoli, S.~Downes, B.~Dutta, F.~G.~Pedro and A.~Westphal,
  JCAP {\bf 1412} (2014) no.12,  030
  doi:10.1088/1475-7516/2014/12/030
  [arXiv:1407.1048 [hep-th]].


\bibitem{Franciolini:2018vbk}
  G.~Franciolini, A.~Kehagias, S.~Matarrese and A.~Riotto,
  JCAP {\bf 1803} (2018) no.03,  016
  doi:10.1088/1475-7516/2018/03/016
  [arXiv:1801.09415 [astro-ph.CO]].


\bibitem{Carr:2009jm}
  B.~J.~Carr, K.~Kohri, Y.~Sendouda and J.~Yokoyama,
  Phys.\ Rev.\ D {\bf 81} (2010) 104019
  doi:10.1103/PhysRevD.81.104019
  [arXiv:0912.5297 [astro-ph.CO]].


\bibitem{Inomata:2017okj}
  K.~Inomata, M.~Kawasaki, K.~Mukaida, Y.~Tada and T.~T.~Yanagida,
  Phys.\ Rev.\ D {\bf 96} (2017) no.4,  043504
  doi:10.1103/PhysRevD.96.043504
  [arXiv:1701.02544 [astro-ph.CO]].


\bibitem{Carr:2017jsz}
  B.~Carr, M.~Raidal, T.~Tenkanen, V.~Vaskonen and H.~Veermäe,
  Phys.\ Rev.\ D {\bf 96} (2017) no.2,  023514
  doi:10.1103/PhysRevD.96.023514
  [arXiv:1705.05567 [astro-ph.CO]].


\bibitem{Bunch:1978yq}
  T.~S.~Bunch and P.~C.~W.~Davies,
  Proc.\ Roy.\ Soc.\ Lond.\ A {\bf 360} (1978) 117.
  doi:10.1098/rspa.1978.0060


\bibitem{Martin:2012pe}
  J.~Martin, H.~Motohashi and T.~Suyama,
  Phys.\ Rev.\ D {\bf 87} (2013) no.2,  023514
  doi:10.1103/PhysRevD.87.023514
  [arXiv:1211.0083 [astro-ph.CO]].


\bibitem{Motohashi:2014ppa}
  H.~Motohashi, A.~A.~Starobinsky and J.~Yokoyama,
  JCAP {\bf 1509} (2015) no.09,  018
  doi:10.1088/1475-7516/2015/09/018
  [arXiv:1411.5021 [astro-ph.CO]].


\bibitem{Kinney:2005vj}
  W.~H.~Kinney,
  Phys.\ Rev.\ D {\bf 72} (2005) 023515
  doi:10.1103/PhysRevD.72.023515
  [gr-qc/0503017].


\bibitem{Tsamis:2003px}
  N.~C.~Tsamis and R.~P.~Woodard,
  Phys.\ Rev.\ D {\bf 69} (2004) 084005
  doi:10.1103/PhysRevD.69.084005
  [astro-ph/0307463].


\bibitem{Namjoo:2012aa}
  M.~H.~Namjoo, H.~Firouzjahi and M.~Sasaki,
  EPL {\bf 101} (2013) no.3,  39001
  doi:10.1209/0295-5075/101/39001
  [arXiv:1210.3692 [astro-ph.CO]].


\bibitem{Gerbino:2016sgw}
  M.~Gerbino, K.~Freese, S.~Vagnozzi, M.~Lattanzi, O.~Mena, E.~Giusarma and S.~Ho,
  Phys.\ Rev.\ D {\bf 95} (2017) no.4,  043512
  doi:10.1103/PhysRevD.95.043512
  [arXiv:1610.08830 [astro-ph.CO]].


\bibitem{Dimopoulos:2005ac}
  S.~Dimopoulos, S.~Kachru, J.~McGreevy and J.~G.~Wacker,
  JCAP {\bf 0808} (2008) 003
  doi:10.1088/1475-7516/2008/08/003
  [hep-th/0507205].


\bibitem{Easther:2005zr}
  R.~Easther and L.~McAllister,
  JCAP {\bf 0605} (2006) 018
  doi:10.1088/1475-7516/2006/05/018
  [hep-th/0512102].


\bibitem{Bond:2006nc}
  J.~R.~Bond, L.~Kofman, S.~Prokushkin and P.~M.~Vaudrevange,
  Phys.\ Rev.\ D {\bf 75} (2007) 123511
  doi:10.1103/PhysRevD.75.123511
  [hep-th/0612197].


\bibitem{Berglund:2009uf}
  P.~Berglund and G.~Ren,
  arXiv:0912.1397 [hep-th].


\bibitem{BlancoPillado:2009nw}
  J.~J.~Blanco-Pillado, D.~Buck, E.~J.~Copeland, M.~Gomez-Reino and N.~J.~Nunes,
  JHEP {\bf 1001} (2010) 081
  doi:10.1007/JHEP01(2010)081
  [arXiv:0906.3711 [hep-th]].


\bibitem{Burgess:2010bz}
  C.~P.~Burgess, M.~Cicoli, M.~Gomez-Reino, F.~Quevedo, G.~Tasinato and I.~Zavala,
  JHEP {\bf 1008} (2010) 045
  doi:10.1007/JHEP08(2010)045
  [arXiv:1005.4840 [hep-th]].


\bibitem{Cicoli:2012cy}
  M.~Cicoli, G.~Tasinato, I.~Zavala, C.~P.~Burgess and F.~Quevedo,
  JCAP {\bf 1205} (2012) 039
  doi:10.1088/1475-7516/2012/05/039
  [arXiv:1202.4580 [hep-th]].


\bibitem{Cicoli:2014sva}
  M.~Cicoli, K.~Dutta and A.~Maharana,
  JCAP {\bf 1408} (2014) 012
  doi:10.1088/1475-7516/2014/08/012
  [arXiv:1401.2579 [hep-th]].


\bibitem{Acharya:2008bk}
  B.~S.~Acharya, P.~Kumar, K.~Bobkov, G.~Kane, J.~Shao and S.~Watson,
  JHEP {\bf 0806} (2008) 064
  doi:10.1088/1126-6708/2008/06/064
  [arXiv:0804.0863 [hep-ph]].


\bibitem{Kane:2011ih}
  G.~Kane, J.~Shao, S.~Watson and H.~B.~Yu,
  JCAP {\bf 1111} (2011) 012
  doi:10.1088/1475-7516/2011/11/012
  [arXiv:1108.5178 [hep-ph]].


\bibitem{Cicoli:2012aq}
  M.~Cicoli, J.~P.~Conlon and F.~Quevedo,
  Phys.\ Rev.\ D {\bf 87} (2013) no.4,  043520
  doi:10.1103/PhysRevD.87.043520
  [arXiv:1208.3562 [hep-ph]].


\bibitem{Higaki:2012ar}
  T.~Higaki and F.~Takahashi,
  JHEP {\bf 1211} (2012) 125
  doi:10.1007/JHEP11(2012)125
  [arXiv:1208.3563 [hep-ph]].


\bibitem{Allahverdi:2013noa}
  R.~Allahverdi, M.~Cicoli, B.~Dutta and K.~Sinha,
  Phys.\ Rev.\ D {\bf 88} (2013) no.9,  095015
  doi:10.1103/PhysRevD.88.095015
  [arXiv:1307.5086 [hep-ph]].


\bibitem{Dutta:2014tya}
  K.~Dutta and A.~Maharana,
  Phys.\ Rev.\ D {\bf 91} (2015) no.4,  043503
  doi:10.1103/PhysRevD.91.043503
  [arXiv:1409.7037 [hep-ph]].


\bibitem{Cicoli:2016olq}
  M.~Cicoli, K.~Dutta, A.~Maharana and F.~Quevedo,
  JCAP {\bf 1608} (2016) no.08,  006
  doi:10.1088/1475-7516/2016/08/006
  [arXiv:1604.08512 [hep-th]].


\bibitem{Allahverdi:2016yws}
  R.~Allahverdi, M.~Cicoli and F.~Muia,
  JHEP {\bf 1606} (2016) 153
  doi:10.1007/JHEP06(2016)153
  [arXiv:1604.03120 [hep-th]].


\bibitem{Pattison:2017mbe}
  C.~Pattison, V.~Vennin, H.~Assadullahi and D.~Wands,
  JCAP {\bf 1710} (2017) no.10,  046
  doi:10.1088/1475-7516/2017/10/046
  [arXiv:1707.00537 [hep-th]].

\end{thebibliography}\endgroup

\end{document}